# Foundational Analysis of Safety Engineering Requirements (SAFER)

Noga Chemo
Tel-Aviv University
Faculty of Engineering
Tel-Aviv, Israel
nogarak@gmail.com

Yaniv Mordecai
Tel-Aviv University
Faculty of Engineering
Tel-Aviv, Israel
ymordecai@tauex.tau.ac.il

Yoram Reich
Tel-Aviv University
Faculty of Engineering
Tel-Aviv, Israel
yoramr@tauex.tau.ac.il

*Abstract*—We introduce a framework for Foundational Analysis of Safety Engineering Requirements (SAFER), a model-driven methodology supported by Generative AI to improve the generation and analysis of safety requirements for complex safety-critical systems. Safety requirements are often specified by multiple stakeholders with uncoordinated objectives, leading to gaps, duplications, and contradictions that jeopardize system safety and compliance. Existing approaches are largely informal and insufficient for addressing these challenges. SAFER enhances Model-Based Systems Engineering (MBSE) by consuming requirement specification models and generating the following results: (1) mapping requirements to system functions, (2) identifying functions with insufficient requirement specifications, (3) detecting duplicate requirements, and (4) identifying contradictions within requirement sets. SAFER provides structured analysis, reporting, and decision support for safety engineers. We demonstrate SAFER on an autonomous drone system, significantly improving the detection of requirement inconsistencies, enhancing both efficiency and reliability of the safety engineering process. We show that Generative AI must be augmented by formal models and queried systematically, to provide meaningful early-stage safety requirement specifications and robust safety architectures.

## TABLE OF CONTENTS

## I. Introduction

Requirements engineering and management are fundamental to product development and are integral to almost all system development models. Despite their importance, existing product development processes often fail to fully reflect the critical role of safety requirements [1]. Safety-critical systems, particularly autonomous systems, involve multiple stakeholders like customers, regulators, and operators, who specify requirements according to their individual priorities [2].

Safety requirements differ from standard system requirements, focusing on accident prevention, risk mitigation, fault tolerance, fail-safe operations, and real-time handling of uncertainties [3]. We distinguish functional safety requirements, which define safe operational behaviors, from probabilistic safety requirements, which quantify acceptable risk levels [4]. A safety architecture maps safety requirements to system functions, integrates protective mechanisms, and mitigates hazards to acceptable levels [5],[6]. Uncoordinated inputs from multiple stakeholders frequently result in contradictions, omissions, and duplications, which can undermine system reliability and safety [7],[8]. Conflicts between protective measure requirements from customers and regulators are all too common. Missing requirements, such as undefined inertial navigation system (INS) drift tolerances, can propagate operational errors, compromise reliability, and impede validation and verification processes [9],[10]. The 2010 Air India Express Flight 812 disaster and other real world accidents underscore the importance of adequate and aligned safety requirement specifications [8].

Model-based systems engineering (MBSE) and model-based safety analysis (MBSA) provide structured methods for documenting requirements and clarifying architectural concepts. Object-Process Methodology (OPM) offers a modeling paradigm that integrates structural and behavioral system aspects into a single framework, improving stakeholder understanding and revealing interdependencies [11],[12]. Manually detecting contradictions and gaps across hundreds of requirements remains unscalable. Duplication and misalignment of safety requirements further complicate the design process and can lead to significant operational risks if not resolved early.

Applying Large Language Models (LLMs) to Model-Based Systems Engineering (MBSE) presents fundamental challenges due to the predominantly graphical nature of MBSE models. MBSE diagrams encode systems semantics with formal syntax: structure, behavior, flow, logic, etc. The diagrams must first be translated into alphanumeric representation for LLM processing. OPM is LLM-ready in that sense, as it offers a rigorous textual modality that is in full semantic sync with the graphical modality. Languages like UML and SysML rely on cryptic data representations like the XML Metadata Interchange (XMI). XMI is a bona-fide representation of the graphical model, but a very complicated one that requires a transformation into a processable structure, which can result in information loss and ambiguity [13]. LLMs are inherently limited in deterministic and consistent analysis of precise behavioral and temporal reasoning, often struggling with synchronization, edge cases, and strict execution semantics. These limitations are particularly critical in safety- and mission-critical MBSE applications, where results must be verifiable, reproducible, and certifiable.

The baseline safety requirements engineering process lacks a unified framework for aligning diverse stakeholder inputs, mapping requirements to system functions, and generating metrics for consistency, completeness, and compliance. Current practices are manual, subjective, and informal, making it difficult to ensure comprehensive and conflict-free requirement sets [14]. Recent advances in Generative Artificial Intelligence (GenAI), Large Language Models (LLMs), and foundational pre-trained token generation offer novel capabilities for analyzing textual requirements and model-based artifacts that may lead to a paradigm shift in systems safety engineering.

We propose a foundational analysis model for safety engineering requirements (SAFER): an automated, model-based methodology and algorithmic process that integrates with MBSE as a source of system requirements and functions, classifies requirements according to system functions, identifies conflicts, similarities, and gaps, and structures the review process for safety architects. SAFER reduces cognitive load, improves analysis efficiency, enhances the reliability of the resulting safety architecture, and informs design decisions. The proposed framework leverages OPM for system modeling and interacts with LLMs via the ChatGPT API to perform comprehensive analyses. Iterative prompt strategies, user audits, and review cycles are employed to optimize the detection of missing, duplicate, and conflicting requirements, differentiating functional and probabilistic safety requirement categories. The SAFER methodology is captured in Figure 1. The **SAFER** algorithm is operated by the **Systems Safety Architect**. It consumes the **System Model** and **generative analysis instructions**. The **System Model** includes both the system architecture – a structural-functional breakdown of the **system** into **sub-systems** and **functionality** into **functions**) and the **stakeholder requirement** sets. The SAFER model generates a report which consists of the following artifacts: a) a **requirement-to-function allocation report**, b) a **requirements classification report**, c) a **conflicting requirement report**, and d) a **duplicate requirement report**.

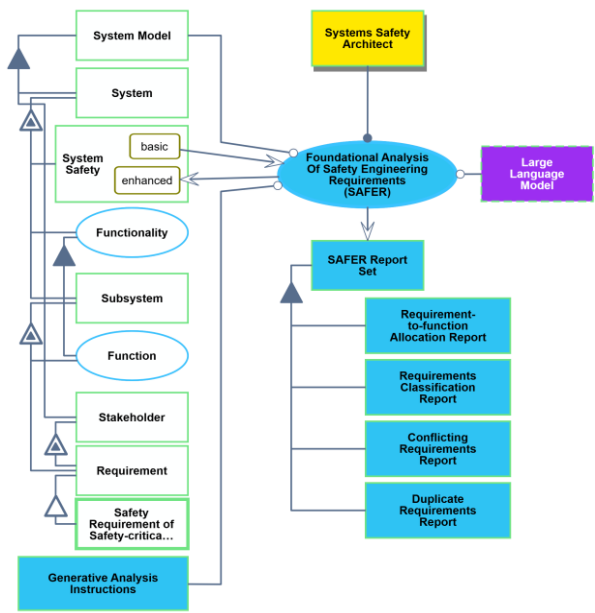

**Figure 1. The SAFER Framework**

We demonstrated SAFER for specifying a safety architecture for an autonomous drone system. SAFER improves classification accuracy, identifies previously overlooked deficiencies, and enables engineers to visualize and address safety function gaps. While the methodology shows promising results, ongoing refinement- including expanding keyword sets, enhancing classification algorithms, and testing across diverse autonomous systems- is required to maximize its utility.

This paper contributes a novel, generative AI-supported methodology for improving safety requirements analysis in complex autonomous systems. By integrating model-based engineering, systematic requirement auditing, and LLM-powered analysis, SAFER enables early detection of conflicts, gaps, and duplications, enhancing the reliability and usability of the resulting safety architecture. The approach demonstrates significant potential for reducing design risks, saving resources, and ensuring compliance with safety standards across a variety of autonomous technologies.

## II. RELATED WORK

### A. Writing safety requirements

Systems engineering is grounded in well-defined requirements. High-quality requirements form the foundation for successful system design, stakeholder alignment, and regulatory compliance. INCOSE defines requirements as precise statements that capture the essential characteristics or constraints of a system, product, or process. These requirements must be unambiguous, consistent, independent, verifiable, and aligned with stakeholder needs. Since requirement definition is the first step of system design, each requirement must satisfy key quality attributes [15]:

**Necessary** – essential for achieving system functionality and regulatory compliance.

**Clear** – expressed simply and concisely to ensure mutual understanding among stakeholders.

**Traceable** – linked both upward to higher-level specifications and downward to detailed ones.

**Verifiable** – testable through methods such as analysis, demonstration, inspection, or testing.

**Complete** – ensuring the system will fully satisfy customer needs while adhering to standards.

### B. Safety Architecture Patterns

Safety Architecture Patterns (SAPs) are reusable micro-architectures designed to simplify and improve the development of safety-critical systems. By capturing expert knowledge in formal model libraries, SAPs provide predefined structures for safety-critical systems design, including redundancy, validated use cases, and safety features.

SAPs combine **informal** attributes (e.g., problem–solution context) with **formal** attributes (e.g., flows, sub-nodes, assumptions) to capture the conceptual and technical aspects of safety-critical systems. This dual representation ensures clarity, consistency, and verifiable guarantees of safety properties. SAPs enable: a) **faster system design** through reuse of validated micro-architectures; b) **automated safety verification** with formal methods (e.g., AltaRica notations, Linear Temporal Logic); c) **efficient allocation** of high-level safety requirements into implementable system designs; and d) **rapid prototyping** of multiple architectures for evaluation and comparison.

*C. Analyzing Requirements*

*1) Identifying Contradictions Between Requirements*

Several methods and frameworks have been proposed to detect contradictions in requirements. ALICE is a system that combines formal logic with Large Language Models (LLMs) to automatically detect contradictions in formal engineering requirements expressed in controlled natural language [16]. ALICE improves both accuracy and recall compared to LLM-only methods and has been demonstrated in industrial contexts like electric bus systems. Accuracy improves thanks to rule-based and deterministic contradiction determinations, rather than probabilistic guesses. *Recall* - an information retrieval performance indicator - is the ratio between the number of correctly identified contradictions and the number of real contradictions. Therefore, when ALICE improves recall, it means ALICE is better at finding contradictions that truly exist, even when they are rare or subtle [15].

Sun et al. explored conflicts between safety and security requirements in cyber-physical systems, particularly in the European car industry [17]. They proposed a framework for detecting contradictions and suggested a co-design approach to integrate safety and security requirements using Maude, a rewriting logic language.

Regulatory contradictions have also been studied. Conflicts between global, regional, and national regulations, such as those affecting agricultural trade, lead to compliance failures and significant economic impacts [18].

Novak and Treytl proposed a Combined Lifecycle Model integrating functional safety (IEC 61508) and system security (Common Criteria) for automation systems, providing conflict resolution rules and assessment metrics across development and operational phases [19].

In air traffic management, Trajectory-Based Conflict Detection (TBCD) and Controller Implemented Separation Standards (CISS) are used to identify and resolve potential trajectory conflicts, emphasizing both automated detection and human oversight [20].

Goal-Oriented Requirements Engineering (GORE) has also been applied to manage conflicting goals across multiple users and requirements, though further automation in conflict detection and decision support is still needed [21].

*2) Identifying Duplicate Requirements*

Duplicate requirements pose challenges in large-scale systems with multiple subsystems and frequent updates. *ReqAligner*, introduced by Rago et al., uses text processing and sequence alignment to identify duplicate use case [22]. Rao et al. addressed duplicate requirements in avionics software using a hybrid model combining *BERT* and *Tree-LSTM* within a Siamese network, integrating semantic and syntactic features for improved detection accuracy and F1 scores [23].

*3) Identifying Missing Requirements*

Incomplete requirements negatively impact system functionality, verification, and development efficiency. Studies have highlighted high rates of missing functionality, constraints, and system interactions in system architecture processes [24], [25].

Bingyang and Harry proposed *UML-based completeness analysis*, using conceptual graphs and inference rules to detect gaps in partially specified models [26].

*The Proxy Viewpoints Model-based Requirements Discovery (PVRD)* methodology supports discovery and management of missing requirements from legacy specifications [27].

*NLP4ReF* is a model for generative analysis of requirement sets, leveraging LLMs and local models to classify subsystems, identify missing requirements, and improve coverage of functional and non- functional requirements [28]. NLP4ReF similarly applied GPT and NLTK for classifying and generating overlooked requirements in IoT projects.

*BERT (Bidirectional Encoder Representations from Transformers)* has been successfully applied to software requirements classification and analysis, achieving high accuracy in both binary and multi-class settings, and enabling detection of contradictions and duplicates [29], [30].

*GloVe (Global Vectors for Word Representation)* is an unsupervised word embedding algorithm that captures semantic relationships through global co-occurrence statistics, improving tasks such as word similarity, analogy, and named entity recognition. Its efficiency and performance make it useful for NLP tasks involving large-scale textual requirement datasets [31].

*D. Model Based Systems Engineering*

*1) OPM*

The *Object-Process Methodology (OPM)* is a holistic framework for system modeling that integrates graphical representation with natural language, offering both rigor and intuitiveness. It distinguishes between *Objectives* (things) and *Processes* (transformations), capturing both structural (static) and behavioral (dynamic) aspects of systems. By supporting *Structure-behavior co-design*, OPM enables safety requirements to be modeled alongside functional and architectural elements, reducing errors and ensuring alignment [32]. It has proven versatile across domains such as defense, aerospace, medicine, software, and molecular biology. In particular, OPM has been applied to analyze *Cyber-Physical Gaps (CPGs)* - deviations between a system's internal model and real-world conditions—which are often root causes of failures and accidents [33].

*2) SysML*

The *Systems Modeling Language (SysML)* is a standardized modeling language for analyzing, designing, and verifying complex systems. It focuses on representing *Functional architecture* using key diagram types: *Use Case, Activity, Block Definition (BDD), Internal Block (IBD), and Sequence Diagrams*. These diagrams capture requirements, structure, and interactions among system components. Traditional SysML model analysis relied on parsing XMI representations, which was hindered by vendor-specific implementations. Recent work, such as by Mordecai et al., introduced algorithms to derive architectures from SysML models and verify them against a *Reference ontology*, ensuring compliance and completeness in system design [13].

*E. Generative Artificial Intelligence (GenAI)*

GenAI enables the creation of synthetic data by learning patterns from real datasets, with applications in *content generation, data augmentation, and systems/requirements engineering (RE)*. Advanced models such as *GANs* and *VAEs* have proven especially effective in the early phases of RE, supporting automated requirements elicitation, analysis, and validation, thereby reducing human error and improving efficiency. A systematic review of 27 studies shows GenAI's strength in these areas but also highlights a research gap in later RE phases, such as requirements management and long-term maintenance. When combined with *Model-Based Systems Engineering (MBSE)*, GenAI offers strong potential for automating and formalizing requirements-related tasks, ensuring consistency in complex systems. However, fully leveraging its potential - particularly for enhancing safety architectures and managing safety requirements in autonomous systems - requires addressing these later-phase challenges.

*1) Bidirectional Encoder Representations from Transformers (BERT)*

BERT (Bidirectional Encoder Representations from Transformers) is a pre-trained language model that enables bidirectional analysis of text, improving various natural language processing tasks [29]. It can perform sentence-level tasks, such as inference and paraphrasing, and token-level tasks, including entity recognition and question answering. In the context of requirements engineering, BERT can help identify contradictory or duplicate requirements by analyzing relationships between sentences. A study applying BERT to software requirements classification achieved over 76% accuracy in both binary and multi-class tasks using the HuggingFace Transformer and BertForSequenceClassification libraries [30]. These results highlight BERT's effectiveness in supporting automated analysis and improving the quality and management of requirement datasets.

*2) Global Vector (GloVe)*

GloVe is an unsupervised learning algorithm that generates distributed word representations by leveraging global word-word co-occurrence statistics from large text corpora [31]. It maps words into a high-dimensional space where semantically similar words are positioned closer together. GloVe combines Global Matrix Factorization and Local Context Window methods to capture both broad statistical patterns and local context, enhancing semantic understanding. The algorithm demonstrates superior performance in tasks such as word similarity, analogy, and named entity recognition, often outperforming other models even with smaller datasets. Its efficiency and robust representation capabilities make GloVe a powerful tool for various natural language processing applications.

*3) Chat GPT*

ChatGPT is an advanced conversational AI model by OpenAI that enables natural language interaction between humans and machines, with significant relevance for safety architecture in autonomous systems [34]. Its capabilities in understanding and generating language can improve decision-making, real-time communication, and risk mitigation in safety-critical contexts. Research highlights its applications across education, healthcare, and customer service, demonstrating benefits such as personalized learning, interactive feedback, and enhanced user engagement [35]. Studies also emphasize ethical considerations, including bias, over-reliance, and the need for oversight when integrating ChatGPT into complex systems. Overall, ChatGPT illustrates the transformative potential of conversational AI in both enhancing user outcomes and supporting safe, resilient autonomous system operations.

The literature demonstrates multiple approaches to identifying contradictions, duplicates, and missing requirements using formal logic, model-based techniques, and advanced NLP methods including LLMs, BERT, and GloVe. Applications range from industrial requirements management, safety-security integration in CPS, air traffic management, and large-scale software systems. Despite progress, full automation, scalability, and integration of these methods into comprehensive safety requirements frameworks remain open research challenges.

III. THE SAFER FRAMEWORK

The objective of this study is to enable early detection of safety requirement issues that could negatively affect later design stages. We present a method for Foundational Analysis of Safety Engineering Requirements (SAFER). We utilize MBSE to capture the system architecture, GenAI to analyze the requirements, and a Python-based model for data integration and analysis.

The SAFER framework is a problem-driven approach to a common systems engineering problem, and aims to improve the performance of the safety architect. We therefore follow meta-methodological guidelines for developing systems engineering methods [36]: building a conceptual model of safety requirements engineering, designing a framework for gap identification, and iteratively developing and testing algorithms for requirement analysis. Performance is measured using defined metrics, with testing conducted on both sample requirements and a real autonomous system case. The approach demonstrates how generative automated reasoning can enhance requirements engineering and safety architecture specification – if applied correctly, carefully, and in coordination with the system architecting process. This paper focuses on and enhances a subset of the scope of the author's research thesis [37].

*A. The Safety Architecture Ontology*

We model the ontology of safety requirements engineering using OPM, as illustrated in Figure 2. This conceptual model defines core safety concepts, classification of requirements as safety requirements, and specialization of safety requirements as probabilistic or functional. Subsystems are defined as safety-critical if they can own safety requirements. This hold for both functional components and ad-hoc safety mechanisms. The notion that functional subsystems may be safety-critical is fundamental since it enables the direct attribution of safety requirements to functional elements of the architecture and makes safety a functional aspect of the system and not only an ad-hoc perspective that lives outside the scope of the system architecture.

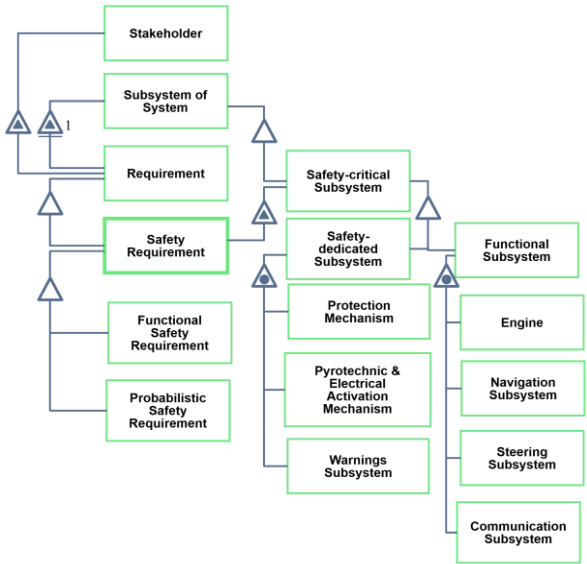

**Figure 2. Safety Architecture Ontology**

*B. The SAFER Protocol*

The SAFER protocol is a structured procedure that helps the safety architect gradually extract critical information from the system model and requirements set, and generate valuable outputs that provide actionable insight. The SAFER protocol is shown in Figure 3. The complete OPL Specification of the SAFER meta-model is provided in Appendix A. We facilitate the generation of outcomes from these processes, by integrating additional supportive processes within SAFER. One such process is initialization, which ultimately leads to an output generation process that contributes to the final stage of the model. SAFER processes include:

a. **Initializing**. setting up the system based on the configuration, and auxiliary services like the LLM client. The configuration includes parameters like input location, LLM connection settings, output location, etc. This process retrieves a list of raw stakeholder requirements and a formal specification of the system architecture model.

**Identifying Subsystems and functions**. This step reads the formal system model's specification. We instruct the SAFER analyzer to ask the LLM to derive the architectural building blocks from the System Architecture Model. The goal of this process is to ensure that the safety architecture is fully aligned with the system architecture, adhering to the core concept of functional safety, whereby safety must be embedded in the system's functionality. This step occurs every time because the architecture may evolve over time and we want to make sure that all the subsystems and functions are covered. SAFER supports importing OPL specifications that represent OPM models and XMI files that represent simple SysML models that include Block Definition Diagram (BDD) specifications (we were able to demonstrate the process on sufficiently simple BDD-based SysML models which contain a limited set of applied notation and element types. We will discuss the differences between the two modeling languages in future publications. SAFER extracts the functional decomposition from the OPL/XMI spec, and uses it as an equivalent architectural basis for subsequent analysis. The extracted core functions are then used throughout the SAFER workflow to classify requirements and to identify missing, duplicate, or contradictory requirements.

b. **Classifying Requirements by Functions and Types**. This step focuses on a) rewriting the requirement to fit the standard requirement writing conventions, b) identifying the subsystem/function to which each requirement pertains, and c) identifying the type of the requirement (functional or probabilistic). The purpose of this step is to cluster requirements by function and type, and to identify unclear requirements which the model is unable to classify into any categorical bucket.

c. **Identifying coverage gaps.** This step aims to ensure that each subsystem/function is sufficiently covered by functional and probabilistic safety requirements and for highlighting subsystems/functions with missing or incomplete safety requirements that could compromise system safety.

d. **Identifying Contradictions**. This step focuses on detecting conflicts among safety requirements from different stakeholders under the same function. We group all requirements by functions, ask the LLM to detect contradictions, and highlight any contradictions in our report.

e. **Identifying Duplications**. This step identifies redundant safety requirements that may lead to ambiguity, inefficiencies or confusion.

f. **Creating output report**. This step generates a report with all the results and provides it to the safety architect for further analysis and action taking.

The SAFER protocol is modular – except for the initialization, any step in the process can be enabled or disabled so that only specific tasks can be achieved. For example, the safety architect can save time and compute by not reading the system architecture again but by using the last available architecture breakdown. In addition, the model can work in Delta mode – if an output has already been generated it will not be regenerated in order to save time and compute resources.

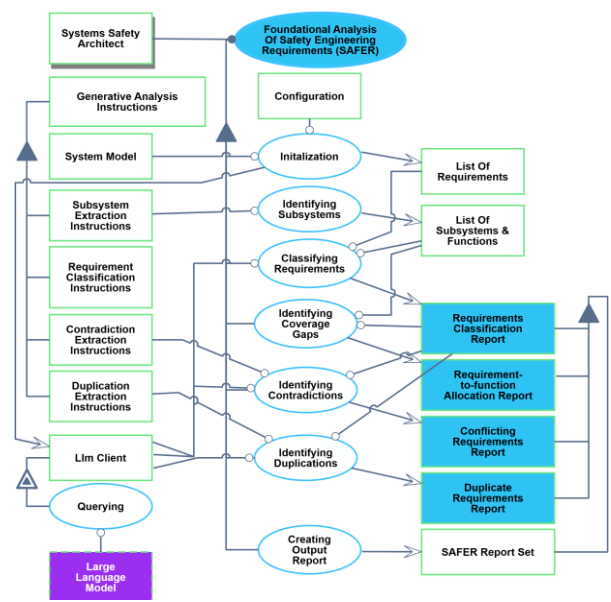

**Figure 3. The SAFER Protocol Meta-Model**

Detected issues should be handled through dedicated problem-solving and solution models, which are then incorporated into the overall system architecture and aligned with the Concept of Operations for practical application. A Stakeholders Model further supports ongoing refinement and adaptation. By systematically addressing requirement deficiencies, SAFER ensures a robust and evolving approach to safety management, with detailed relationships.

*C. SAFER Performance Metrics*

The success of SAFER depends on meeting several key functional criteria. SAFER must accurately obtain system core functions from a formal model and correctly allocate safety requirements to them. To ensure reliability, SAFER needs to detect missing, duplicate, and contradictory requirements, thereby improving completeness, clarity, and consistency. The key performance metrics for those criteria are:

- Subsystem identification: >90% accuracy.
- Requirements classification: >80% accuracy.
- Identifying Duplicates: >80%.
- Contradictory/Conflicting Requirements: >80% accuracy.
- Classification Stability/consistency: >80% consistent accuracy across multiple runs.

*D. LLM-Based Requirements Analysis and Classification*

SAFER leverages a Large Language Model (LLM) to perform several tasks, including the identification of system functions, classification of requirements by system functions, distinguishing between *functional* and *probabilistic* requirements, detecting contradictions, and detecting duplications. To achieve reliable results, the process is structured around three pillars:

1. Prompt Composition – Constructing rich, informative prompts with clear semantic and syntactic instructions, supported by reference dictionaries for precision.

2. Batching of Requirements – Dividing large sets into configurable batches, improving speed, precision, and consistency while ensuring classification is unbiased by prior knowledge. This practice also provides for important cross-contextualization – a phenomenon that causes the LLM to consider all the tokens in the input while acting on each item.

3. Programmatic API Access – Using ChatGPT's API for automated classification, parsing, and mapping of responses to the input requirements.

*E. Identifying subsystems and functions*

The objective is to correctly extract all the subsystems and functions from system architecture model. The output should be provided as a JSON structure which provides the system, subsystem, and function as a hierarchical structure, as illustrated below. This architecture representation data structure captures both the structural decomposition (system-subsystem) and functional decomposition (functionality-function) and thus allows the designer to work both on structural features (modules, components, etc.) and functional ones (behaviors, processes, interactions, etc.). The complete prompt is provided in Appendix B.

```
{
 "Drone": {
  "NAV": "Drone/Navigation/Navigating",
  "EN": "Drone/Engine/Power Generating",
  "TD": "Drone/Communication/Data Transmission",
  "RD": "Drone/Communication/Data Receiving",
  "PEA": "Drone/Mission System/Pyrotechnic and Electrical Activation",
  "_OF_": "Other Function"
 },
 "Environment": {
  "AF": "AirFlow"
 },
 "Operator": {
  "CTRL": "Controlling",
  "MNTR": "Monitoring"
 }
}
```

*F. Finding Missing Requirements*

The objective of "Finding Missing Requirements" is to identify core functions of an autonomous system that lacks necessary requirements, ensuring completeness in functional and probabilistic aspects. The steps are:

1. Consolidate all requirements into one list

2. Loading requirements list and core system functions (as identified in Phase 1)

3. Building a Function-Type Matrix, with which core functions are evaluated

4. Asking the API to classify each requirement to the right function and classify it to Functional or Probabilistic.

The complete prompt is provided in Appendix C.

*G. Finding Duplicate Requirements*

The objective of *"Finding Duplicate Requirements"* process, is to detect and address redundancy in the requirements list, ensuring that each requirement is unique and non-redundant. The steps are: 1. Clustering Requirements by Function, and 2. Duplicate Detection Request. The unified list of all the requirements for the specific function is submitted to the LLM's API with a specific request to identify and report duplicate requirements. The LLM uses its capabilities to analyze the text and identify any repetitive entries in the list.

*H. Finding Contradicting Requirements*

The objective of *"Finding Contradicting Requirements"* is to identify requirements that conflict with each other, which may lead to inconsistencies or operational issues within the autonomous system. The steps are: 1. Clustering Requirements by Function, and 2. Conflict Detection Request.

The consolidated list without duplicates is sent to the LLM's API with a request to identify conflicting requirements. The LLM analyzes the list to detect any contradictions or conflicts between the requirements based on their content and context.

*I. Implementation with Python*

SAFER is a python package which provides several key services that enable seamless generative analysis[1]. The model consumes a parameters file that allows the user to specify multiple analytical tasks, to be executed in a sequential order. Each task block includes the following parameters:

```
"b_classify_requirements":{
 "type":"GENERATIVE_ANALYSIS_TASK",
 "run":true,
 "delta": true,
 "project_dir": "C:/models/Drone",
 "readme": "readme.txt",
 "input_file":
"B_Requirements/input/safety_requirements.csv",
 "dataset_name": "Drone Safety Requirements",
 "dataset_id_column": "ReqID",
 "dataset_columns": ["Requirements"],
 "result_columns":["Function","Type",
"Confidence", "System_Requirement"],
 "instructions":"B_Requirements/instructions.txt",
 "resources":
"B_Requirements/data_dictionary.json",
 "output_path": "B_Requirements/results",
 "chunk_size":10,
 "max_items": -1,
 "execute":true,
 "analyze":true,
 "analysis_function":"analyze_requirement_completet
ness",
 "verbose":true
}
```

---

[1] *A link to the final version of the code will be provided in the camera ready version.*

The SAFER script is essentially a task runner for generative analysis workflows. For each task in the configuration, it combines defaults with task-specific parameters, validates them, and then prepares the working environment: sets the project directory, reads a README if one exists, assigns a version tag (defaulting to today's date), and defines input and output file paths. It also reads in an API key, instructions for the analysis, and contextual resources (which it tries to convert to JSON, or falls back to raw text). The resources and dataset become the context for the generative process.

Once inputs are ready, the script loads the dataset, breaks it into chunks if needed, and—if the task's execute flag is true—begins execution. The execution function handles the heavy lifting: assembling prompts from instructions, resources, and data, sending them to an LLM service, and writing results to versioned output files. After execution, the script has a provision for parsing the results into a tabular structure, matching the results with the inputs, and running a task-specific analysis function on the results. In short, the script automates the setup, resource management, and orchestration of multiple generative AI analysis tasks. In fact, it is almost agnostic to the content that it consumes or generates, with the exception of the dedicated result analysis functionality, which requires domain-specific and specialized constructs and processes for analysis and insight generation.

## IV. APPLYING SAFER IN DRONE SAFETY ENGINEERING

We applied SAFER to autonomous drone design, and demonstrated its role in generating insights and improving design results. The objective is to show how SAFER and similar frameworks can be studied, evaluated, and refined.

Drones create various opportunities but also raise safety concerns and operational challenges that must be addressed for their safe integration. Key safety aspects include cybersecurity vulnerabilities, which expose drones to hacking and unauthorized control; physical safety risks such as crashes or collisions; privacy concerns from surveillance capabilities; environmental risks in harsh weather; and issues with the human–machine interface and operator training that can lead to errors. The main challenges include ensuring technological robustness and resilience, closing regulatory and legal gaps, managing operational complexity and coordination with other systems, implementing rigorous risk assessment and safety practices, and balancing growing autonomy with the need for human oversight in critical environments.

The case study focuses on core components typical to autonomous systems: a frame, BLDC motors, propellers, motor controllers, battery, transmitter/receiver, and a flight controller, which compose the system's architecture and operational context. Each component belongs to a systemic function. Figure 4 illustrates the drone's safety architecture, focusing on the safety-critical systems and functions.

### A. Application of SAFER

The application of SAFER was carried out in two phases: **Exploration** and **Exploitation**. In the *Exploration phase*, a small set of 42 requirements (Req Set A) was analyzed to gain initial insights into the model's behavior, input–output dynamics, and the LLM's capabilities. Although results were expected to be limited, this phase provided critical understanding for refining the process. In the *Exploitation phase*, a larger set of 110 requirements (Req Set B) was analyzed using the insights gained earlier. This phase systematically addressed key SAFER tasks—sufficiency, duplication detection, and contradiction detection—while delivering more robust and reliable results. This application relies only on textual requirements that we collected and provided to SAFER as a raw input. Future research will focus on enhancements such as model-based requirements engineering and formal requirement specifications.

### 1) Phase 1: Exploration

We began the exploration with an initial set of 42 safety requirements. We classified each requirement manually under one of the system's safety-critical functions, in order to evaluate the accuracy of the LLM's response. Those requirements are intentionally "messy", to imitate the common situation, in which stakeholder requirements are provided in raw, often unprofessional language, without care for the appropriateness, accuracy, and articulation of the requirements.

We expected each requirement to be categorized into a defined safety function: navigation, steering, supervision, etc. We also expected that each requirement would have a classification of whether it is a functional or probabilistic requirement.

We ran this exercise again, this time with an incremental step of 1 requirement per iteration, to obtain better understanding of the behavior of the model with every incremental addition of requirements. We witnessed the same behaviors.

We evaluated the consistency of the LLM's classification by matching it with the expected classification, and observed that the consistency achieved was 71.42%. The consistency was insufficient. Therefore, it was imperative to refine the model.

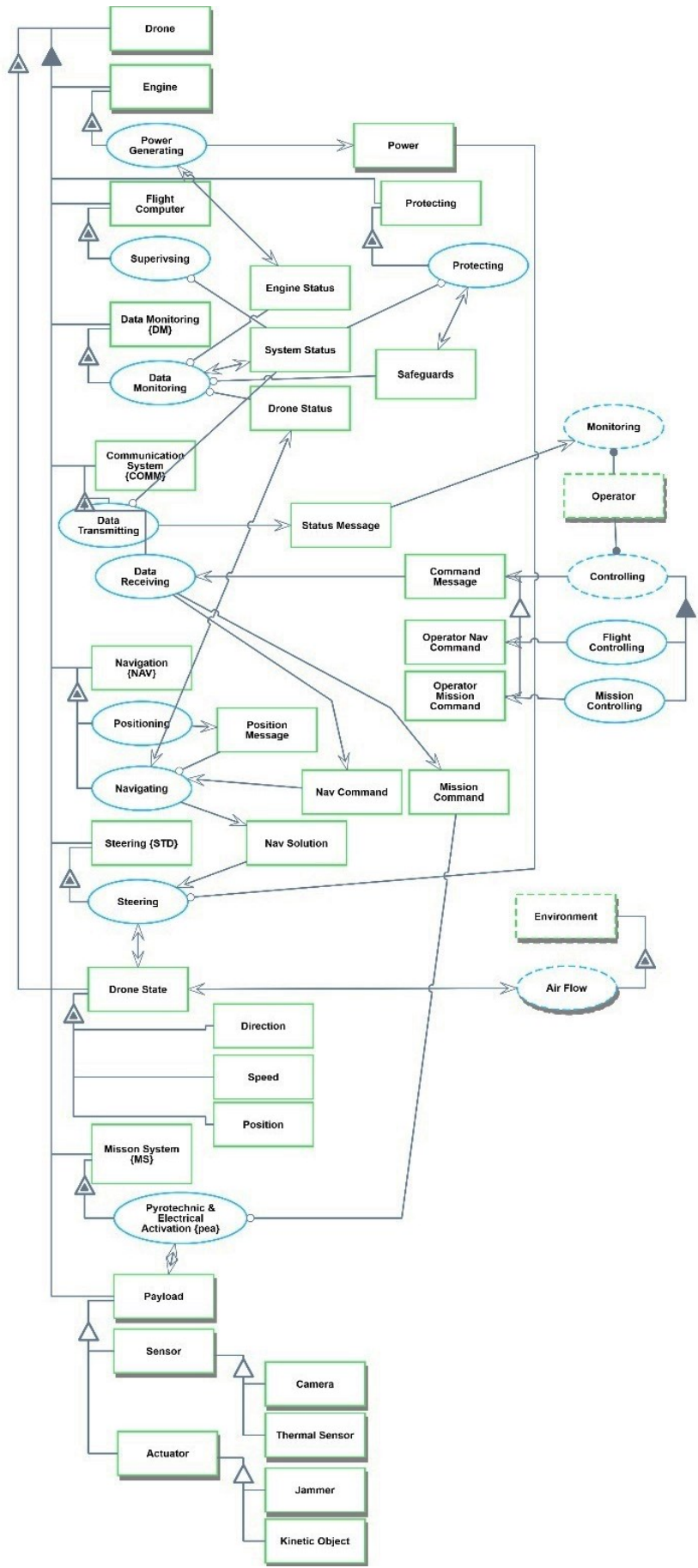

**Figure 4. High-Level Drone Safety Architecture**

We believe that classification inconsistency occurs following a comparison of pairs of requirements. This comparison classifies each requirement's function, with reference to the classifications of the neighboring requirements. LLMs learn from the context provided to them. A short or missing context may be insufficient, but an overly long context may be more confusing and misleading to the LLM. The classification prompt was refined to improve accuracy and transparency. First, the LLM was instructed to classify each requirement independently of others. Second, a dummy function "_OF_" was introduced to handle ambiguous cases. Third, the LLM was required to explain its reasoning for each classification, enhancing explainability. Finally, it was asked to provide a confidence level for each result to support evaluation of reliability.

We ran SAFER with the additional instructions and 18 requirements were classified under the "_OF_" category. In deciphering the reasons, it seems that a badly-phrased requirement was a common reason. When the requirements were refined and clearly redefined, the outcomes showed significant improvement and reached 81.82% accuracy.

To improve the classification process again, we performed several actions: a) we added an "information security" function, b) we clarified the wording of the 18 misclassified requirements, and c) we removed mission requirements from the input set.

We conducted another run of SAFER. This iteration revealed that, of the 18 requirements previously classified as "_OF_", only 16 requirements were categorized under the "_OF_" designation this run. Following this refinement, an increased stability in requirements mapping to functions was noted in each iteration, yielding 82.72% accuracy.

*2) Phase 2A: Assessing Requirement Set Sufficiency*

We applied SAFER to classify and aggregate the 110 requirements and then arrange them in a Function-Type matrix to evaluate sufficiency. A sample of the results is shown in Appendix D. **Sufficiency was defined as having at least three functional requirements and one probabilistic requirement per system function.** This choice stems from several reasons: a. One input requirement, One processing requirement, and One output requirement for each core system function; and b. Rule of thumb based on field experience: each function usually has at least three functional requirements. The expectation for at least one probabilistic requirement is a proactive approach encouraging stakeholders to apply stochastic thinking, since it is actually not always common to see probabilistic requirements, mainly for safety-critical functions that are not safety-specific (e.g., engines, flight computer, etc.). Applying this analysis to Req Set B showed that while most functions had sufficient functional requirements, probabilistic requirements were lacking: 4 of 9 functions had none, and 4 had only the bare minimum. Additionally, unclassified requirements did not contribute to addressing the probabilistic gap, highlighting a critical shortfall in this area, as shown in Table 1.

The output indicates which functions are sufficiently specified and which remain partially specified, in accordance with the specified requirements. This output empowers the system architect to identify the functions for which additional requirements are necessary.

*Table 1: Function-Type Requirement Set Sufficiency Matrix*

| Type | Function | N_Reqs FUNC | PROB | Result |
|---|---|---|---|---|
| 0 | DM | 18 | 2 | Complete |
| 1 | EN | 9 | 1 | Complete |
| 2 | NAV | 12 | 0 | Missing |
| 3 | PEA | 14 | 0 | Missing |
| 4 | PTC | 6 | 1 | Complete |
| 5 | RD | 3 | 0 | Missing |
| 6 | STR | 2 | 1 | Missing |
| 7 | SUP | 15 | 1 | Complete |
| 8 | TD | 6 | 0 | Missing |
| 9 | OF | 16 | 0 | Missing |

*3) Phase 2B: Identifying Duplicate Requirements*

In this phase we increased the size of the requirements set to 110 requirements and conducted a more comprehensive analysis. To facilitate the identification of duplicate requirements, we sent the 110 requirements through the chatGPT API in functional chunks, i.e. all the requirements of each function were sent together in a single call, with the following prompt:

```
For all the requirements in the list, mark
the duplicate requirements.
```

We expected that SAFER will mark the duplicate requirements. The results show two phenomena: designating non-duplicate requirements as duplicates (false positive, or FP) and failing to mark actual duplicate requirements as such (false negative or FN).

*a) Designating False Positive requirements*

Non-duplicate requirements can be mistakenly identified as duplicates due to limitations in the language model's textual processing. The model relies heavily on surface-level similarity, often missing nuanced differences in context or intent, which is critical for safety requirements. Factors contributing to misclassification include low similarity thresholds, common phrasing in safety standards, ambiguity, redundancy, and potential overfitting to training data. As a result, requirements that are functionally distinct may be incorrectly flagged as duplicates.

*b) Designating False Positive requirements*

Misclassification of duplicate requirements can arise from several factors: insufficient context in prompts, causing the model to misinterpret poorly defined requirements; simplistic similarity scoring that ignores semantic meaning; limitations of the ChatGPT API in handling complex or domain-specific comparisons; inherent ambiguity in some safety requirements; and an inappropriate threshold for determining duplicates, which can either overlook true duplicates or falsely flag non-duplicates. Addressing these issues is key to improving accuracy in duplicate detection.

The prompt was updated to focus on defining system functions clearly, so that redundancy among requirements could be correctly interpreted as identical specifications for the same function, improving duplicate detection accuracy.

```
For all the requirements in the list, mark
the duplicate requirements.

If two requirements are similar but refer to
two different functions it is not considered
duplicate.
```

In this analysis, SAFER detected only 3 out of 8 known duplicate requirement pairs (37.5% success). The low detection rate is attributed to duplicate requirements being classified under different categories, preventing the algorithm from recognizing them as duplicates. For example, similar monitoring requirements were assigned to separate functions (engine vs. data monitoring), suggesting that aligning categories could improve duplicate identification. Therefore, to increase the compatibility level, we enhanced the prompt:

```
For all the requirements in the list, mark
the duplicate requirements.

If two requirements are similar but refer to
two different functions it means that they
are complementary.

If two requirements are similar and refer to
the same function it means that they are
duplicate.

If one of the requirements refers to the
function "_OF_" ("Other Function") it could
mean that the requirement refers to a system-
level functionality or to each one of the
functions. In this the specific function's
requirement might be a refinement of the top
level requirement.
```

This time, SAFER detected 7 out of 8 duplicate pairs, resulting in a compatibility rate of 87.5%. This level of accuracy is considered satisfactory, indicating that the refinement of the prompt effectively addressed redundancy in the requirements set.

*4) Phase 2C: Identifying Contradicting Requirements*

We facilitated the identification of contradicting requirements, by submitting the requirements grouped by functions, and the following prompt:

```
For all the requirements in the list, mark
the contradicting requirements.
```

SAFER successfully detected 7 out of 9 known contradictory requirement pairs, achieving a 77.8% accuracy. Missed contradictions may result from ambiguities in requirement phrasing, complex interdependencies between requirements, or limitations in the model's contextual understanding of domain-specific nuances. Despite these challenges, results indicate that SAFER is generally effective at identifying contradictions within the requirement set.

SAFER misclassified one non-contradictory requirement pair as contradictory. This may result from an overly conservative analysis aimed at ensuring safety, insufficient contextual analysis that overlooks conditional or complementary relationships, or potential errors in the classification logic. Such misclassifications highlight areas for refinement in SAFER's conflict detection capabilities.

## V. DISCUSSION & CONCLUSION

This study addresses challenges in safety requirements for autonomous systems, such as inconsistencies, duplications, and omissions, which can negatively impact system design. The SAFER model was developed using Model-Based Systems Engineering (MBSE) and Generative AI to enable early identification and resolution of these issues, improving the reliability and regulatory compliance of safety architectures. SAFER involves modeling safety functions and requirements, establishing a theoretical framework for gap identification, designing an analysis mechanism, iteratively developing and testing an algorithm, and applying it to a realistic autonomous system case study. The application of SAFER to an autonomous drone system demonstrated the model's effectiveness through two phases: an exploratory phase analyzing 42 requirements to calibrate the model, and an exploitation phase analyzing 110 requirements to systematically assess sufficiency, duplicates, and contradictions. SAFER has performed well, and provided results within seconds for each of the prompts, saving significant effort and sparing precious designer time for focusing on corner cases and anomalies. Overall, SAFER provides a structured, systematic approach for analyzing and refining safety requirements to strengthen autonomous system safety architectures. SAFER provides a robust foundation for further extension and integration with additional tools and artifacts within the system architecture domain.

*A. Advantages of SAFER*

Running SAFER repeatedly showed improved results. Adding 68 requirements to the requirement set (a 150% increase) showed a significant improvement in identifying missing requirements. Improving the prompt also contributed to model performance, highlighting the importance of sound instruction specification. This ability to iterate in requirements analysis has proved that repeated runs, with precision in guidance, improve the results of the run. Running the model over and over again was possible through robust configuration. Thus, SAFER provides a **systematic, repeatable, and scalable approach** for analyzing safety requirements, ensuring that all requirements are consistently evaluated against one another. This structured methodology minimizes the risk of oversight and helps identify conflicts or contradictions that might otherwise go unnoticed, which is crucial for complex autonomous systems.

SAFER's **robust textual analysis** leverages advanced natural language processing techniques, enabling the detection of contradictions, ambiguities, and overlapping requirements. By analyzing requirements based on both semantic meaning and context, SAFER can uncover issues that are not immediately apparent from surface-level wording.

By identifying contradictions **early in the design process**, SAFER significantly **enhances safety assurance**, allowing engineers to refine and harmonize safety requirements before development. This proactive approach reduces the likelihood of safety issues and ensures that the system design is both comprehensive and coherent.

SAFER also improves **communication among stakeholders**, as its detailed analyses provide concrete evidence of conflicts and discrepancies. This clarity facilitates informed discussions, collaborative problem-solving, and iterative refinement of safety protocols.

Furthermore, SAFER ensures **documentation and traceability**, maintaining a clear record of how each requirement was evaluated and how conflicts were resolved. This is particularly valuable for regulatory compliance, audits, and future reference in system upgrades or modifications.

SAFER saves the architect time, effort, and costly work hours, and reduces the detection margin for human inspection to a reasonable size, preventing delayed detection of issues and preventing downstream system functionality and safety.

*B. Recommended Model Improvements*

  *1) Improving Missing Requirements Detection*

  When certain system functions still lack full requirement coverage, highlighting and prioritizing areas for further design elaboration may be more useful and actionable than simply listing the coverage gaps.

  *2) Improving Duplicate Requirements Detection*

  SAFER's duplicate detection can be enhanced by addressing redundancy within and across categories, distinguishing complementary requirements, and consolidating similar items for clarity. Refining the prompt to focus on intent and context, implementing semantic analysis, adjusting similarity thresholds, and providing illustrative examples can improve accuracy. These improvements aim to reduce false positives and increase recall/accuracy.

  *3) Improving Requirements Contradiction Process*

  SAFER has shown promising results in detecting contradictions in complex and ambiguous requirements. Further refinement of algorithmic logic and contextual understanding, combined with adherence to IAEA Safety Guide guidelines, can improve conflict detection accuracy and compliance. Future work should test SAFER across different autonomous systems (e.g., cars, missiles, UAVs) and expand keywords and derivatives to identify additional potential conflicts, enabling broader validation and model development.

*C. Limitations of the Current Study*

While SAFER significantly improves the identification of safety requirement issues, it has several limitations. Its effectiveness depends on keyword matching, meaning requirements that do not align with predefined keywords may go undetected. The final identification of contradictions still relies on the architect's judgment, introducing potential human error. Initial evaluations revealed classification inconsistencies, with some functions still lacking complete specifications, and the model may have limited generalizability across different autonomous systems. SAFER has been tested for a safety system model with an input size of 110 safety requirements. We have seen that as the number of requirements fed into the model increased, the level of significance increased. We performed dozens of runs of the model and examined the stabilization of the answers from the model, by using a growing batch size from 1 through 40, which trades off with response time and input token limitation. The set of 110 requirements is considered medium sized, and sufficient for a functional demonstration on a relatively familiar autonomous drone system. In future studies, we will examine: a. Stability of the answer in cycles with a variable batch size in order to find an optimal batch size without compromising the accuracy and consistency of the model (we set the model temperature to zero to achieve a deterministic answer as much as possible), b. Providing the model with expert-validated classifications that the model provided in previous runs, so that the model can correct mistakes it made and learn from them for subsequent runs and cases, c. Testing SAFER on a project with thousands of mission requirements in order to assess SAFER on large-scale projects, d. Investigating anomalies observed during execution with ChatGPT, as they could inform future refinements, and e. Extending the framework, which currently focuses on pairwise requirement analysis, to higher-order inter-requirement contradiction searches.

The limitations highlight the need for ongoing refinement of SAFER, particularly regarding keyword coverage, contextual understanding, and algorithm logic. Future research could explore higher-order interactions among requirements, improving robustness for complex systems. Testing the model on diverse autonomous systems, including vehicles, missiles, UAVs, and household robotics, will help assess and enhance SAFER's applicability. Expanding prompt options and keyword fine-tuning could further empower architects to identify contradictions and gaps more effectively.

Despite these limitations, SAFER remains a valuable tool for requirements management, providing a structured, AI-assisted framework to detect gaps, duplications, and contradictions. Its systematic approach supports decision-making, enhances the reliability of safety requirements, and strengthens overall safety architecture in autonomous systems. Continuous improvements are expected to increase compliance rates, accuracy, and generalizability across various domains.

Although SAFER was developed and evaluated in the context of safety-critical autonomous systems, there is no inherent limitation that prevents its application to other engineering domains. The methodology is fundamentally driven by the explicit definition of core system functions and the structured allocation of functional and probabilistic requirements to those functions. In principle, SAFER can be extended to general engineering or mission-oriented systems by redefining the functional decomposition (e.g., navigation, guidance, control, steering, mission logic, or actuation) and the analytical ontology. Under these conditions, SAFER can classify requirements, assess functional coverage, and identify missing, duplicate, or conflicting requirements in a manner analogous to its safety-focused application. However, meaningful results depend on domain-specific expertise, appropriate function definitions, and tailored requirement categories, suggesting that successful transfer beyond safety-critical systems will require careful adaptation rather than as-is model reuse.

As for human-machine-balance, as safety requirements can potentially cause a catastrophic event and harm to human life, human review of the output is strongly advised. The model facilitates the process, saves time, and assists the systems architect – it is not intended to replace the human in the loop. Human validation is useful as feedback to the LLM regarding previously suggested classifications. Therefore, the right balance is to use both a model and human input for analyzing the safety requirements and safety architecture.

In future studies we aim to hold a comparative analysis of several LLMs (ChatGPT, Claude, Perplexity, Gemini) and establish a baseline test for safety requirements verification that may provide a common standard for similar analyses by other researchers. Professionals who reviewed this work suggested extending its applicability to general system requirements. While this is possible, we conclude based on our experience with SAFER that it will be crucial to deal with various categories of requirements with ad-hoc treatment and expertise, as it was critical for achieving meaningful and actionable results in the safety domain.

## BIOGRAPHY

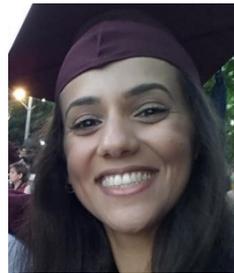

*Noga Chemo is a senior aerospace systems engineer. She received her M.Sc. in Systems Engineering (2024) from Tel Aviv University. She holds a B.Sc. in Industrial Engineering from Holon Institute of Technology (2013, Summa Cum Laude). Her research interests include Safety Engineering, Applications of AI for Systems Analysis, and applications of model-based systems engineering to the design of resilient safety-critical systems.*

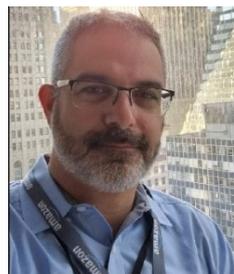

*Yaniv Mordecai is an adjunct professor at Tel-Aviv University (TAU), Tel-Aviv, Israel, and a Senior Research Scientist at Amazon, Bellevue, Washington, USA. He was a Technion-MIT Post-Doctoral Fellow at MIT's Engineering Systems Laboratory (2019-21), He holds a Ph.D. in systems engineering from Technion – Israel Institute of Technology, Israel (2016), and M.Sc. (2010, cum laude) and B.Sc. (2002) degrees in industrial engineering & management from TAU, Israel. His research interests include model-based systems engineering, model analytics, cybernetics, interoperable systems, risk and decision analysis, and operations research. Dr. Mordecai is a senior member of IEEE, member of INCOSE, Board Member of the Israeli Society for Systems Engineering – INCOSE_IL, member of INFORMS, and Duly Elected Member of Signa Xi. He won two international doctoral research awards and several "Best Paper" Awards.*

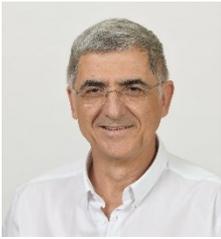

**Yoram Reich** received the B.Sc. (summa cum laude) and M.Sc. (magna cum laude) degrees in mechanical engineering from Tel-Aviv University (TAU), in 1980 and 1984, respectively, and the Ph.D. degree in civil engineering from Carnegie Mellon University, in 1991. He is currently a Chaired Professor of Engineering Design and Systems Engineering with the Faculty of Engineering, TAU, Tel Aviv, Israel. He has co-authored more than 250 papers and four books. His research interests include product design methods and theories, systems engineering, and innovation. He is an Editor-in-Chief for Research in Engineering Design, a Senior Editor for Design Science, and a member of the editorial board of seven other journals. He is a Fellow of the Design Society, a Fellow of the Design Research Society, an Honorary Fellow of INCOSE-IL, and a member of INCOSE. He is also a co-founder and head of the TAU Systems Engineering Research Initiative.

APPENDIX A. OPL SPECIFICATION OF THE SAFER MODEL

1. **System** is an informatical and systemic object.
2. **Stakeholder** is an informatical and systemic object.
3. **Requirement** of **Stakeholder** and **Subsystem** is an informatical and systemic object.
4. **Systems Safety Architect** is a physical and systemic object.
5. **Safety Requirement of Safety-critical Subsystem** is an informatical and systemic object.
6. **SAFER Report Set** is an informatical and systemic object.
7. **Subsystem** of **System** is an informatical and systemic object.
8. **System Model** is an informatical and systemic object.
9. **Generative Analysis Instructions** is an informatical and systemic object.
10. **Requirement-to-function Allocation Report** is an informatical and systemic object.
11. **Requirements Classification Report** is an informatical and systemic object.
12. **Conflicting Requirements Report** is an informatical and systemic object.
13. **Duplicate Requirements Report** is an informatical and systemic object.
14. **System Safety** of **System** is an informatical and systemic object.
15. **System Safety** of **System** can be **basic** or **enhanced**.
16. **Large Language Model** is an informatical and environmental object.
17. **Stakeholder** exhibits **Requirement**.
18. **Safety Requirement of Safety-critical Subsystem** is a **Requirement**.
19. **System** exhibits **Subsystem** and **System Safety**, as well as **Functionality**.
20. **Subsystem** exhibits **Requirement**, as well as **Function**.
21. **System Model** consists of **Stakeholder** and **System**.
22. **SAFER Report Set** consists of **Conflicting Requirements Report**, **Duplicate Requirements Report**, **Requirement-to-function Allocation Report**, and **Requirements Classification Report**.
23. **Functionality** consists of **Function**.
24. **Foundational Analysis Of Safety Engineering Requirements** is an informatical and systemic process.
25. **Foundational Analysis Of Safety Engineering Requirements** changes **System Safety** of **System** from **basic** to **enhanced**.
26. **Systems Safety Architect** handles **Foundational Analysis Of Safety Engineering Requirements**.
27. **Foundational Analysis Of Safety Engineering Requirements** requires **Generative Analysis Instructions**, **Large Language Model**, and **System Model**.
28. **Foundational Analysis Of Safety Engineering Requirements** yields **SAFER Report Set**.
29. **Function** of **Subsystem** is an informatical and systemic process.
30. **Functionality** of **System** is an informatical and systemic process.
31. **Safety Requirement** of **Safety-critical Subsystem** from SD specialization-unfolds in SD1 into **Functional Safety Requirement** and **Probabilistic Safety Requirement**.
32. **Safety Requirement** of **Safety-critical Subsystem** is an informatical and systemic object.
33. **Requirement** of **Stakeholder** and **Subsystem of System** is an informatical and systemic object.
34. **Functional Safety Requirement** is an informatical and systemic object.
35. **Probabilistic Safety Requirement** is an informatical and systemic object.
36. **Subsystem of System** is an informatical and systemic object.
37. **Stakeholder** is an informatical and systemic object.
38. **Safety-critical Subsystem** is an informatical and systemic object.
39. **Functional Subsystem** is an informatical and systemic object.
40. **Safety-dedicated Subsystem** is an informatical and systemic object.
41. **Engine** is an informatical and systemic object.
42. **Navigation Subsystem** is an informatical and systemic object.
43. **Steering Subsystem** is an informatical and systemic object.
44. **Communication Subsystem** is an informatical and systemic object.
45. **Protection Mechanism** is an informatical and systemic object.
46. **Pyrotechnic & Electrical Activation Mechanism** is an informatical and systemic object.
47. **Warnings Subsystem** is an informatical and systemic object.
48. **Safety Requirement** is a **Requirement**.
49. **Functional Safety Requirement** and **Probabilistic Safety Requirement** are **Safety Requirements**.
50. **Stakeholder** exhibits **Requirement**.
51. **Subsystem of System** exhibits **Requirement** and one more operation.
52. **Safety-critical Subsystem** is a **Subsystem of System**.
53. **Safety-critical Subsystem** exhibits **Safety Requirement**.
54. **Functional Subsystem** and **Safety-dedicated Subsystem** are **Safety-critical Subsystems**.
55. **Communication Subsystem**, **Engine**, **Navigation Subsystem**, and **Steering Subsystem** are instances of **Functional Subsystem**.
56. **Protection Mechanism**, **Pyrotechnic & Electrical Activation Mechanism**, and **Warnings Subsystem** are instances of **Safety-dedicated Subsystem**.
57. **Foundational Analysis Of Safety Engineering Requirements** from SD part-unfolds in SD2 into **Classifying Requirements**, **Creating Output Report**, **Identifying Contradictions**, **Identifying Coverage Gaps**, **Identifying Duplications**, and **Initalization**.
58. **Systems Safety Architect** is a physical and systemic object.
59. **Generative Analysis Instructions** is an informatical and systemic object.
60. **System Model** is an informatical and systemic object.
61. **SAFER Report Set** is an informatical and systemic object.
62. **List Of Requirements** is an informatical and systemic object.
63. **List Of Subsystems & Functions** is an informatical and systemic object.
64. **Large Language Model** is an informatical and environmental object.
65. **Subsystem Extraction Instructions** is an informatical and systemic object.
66. **Requirement-to-function Allocation Report** is an informatical and systemic object.
67. **Requirements Classification Report** is an informatical and systemic object.
68. **Conflicting Requirements Report** is an informatical and systemic object.
69. **Duplicate Requirements Report** is an informatical and systemic object.
70. **Contradiction Extraction Instructions** is an informatical and systemic object.
71. **Duplication Extraction Instructions** is an informatical and systemic object.
72. **Llm Client** is an informatical and systemic object.
73. **Configuration** is an informatical and systemic object.
74. **Requirement Classification Instructions** is an informatical and systemic object.
75. **Foundational Analysis Of Safety Engineering Requirements** consists of **Classifying Requirements**, **Creating Output Report**, **Identifying Contradictions**, **Identifying Coverage Gaps**, **Identifying Duplications**, and **Initalization**.

76. **Generative Analysis Instructions** consists of **Contradiction Extraction Instructions**, **Duplication Extraction Instructions**, **Requirement Classification Instructions**, and **Subsystem Extraction Instructions**.
77. **SAFER Report Set** consists of **Conflicting Requirements Report**, **Duplicate Requirements Report**, **Requirement-to-function Allocation Report**, and **Requirements Classification Report**.
78. **Llm Client** exhibits **Querying**.
79. **Foundational Analysis Of Safety Engineering Requirements** is an informatical and systemic process.
80. **Systems Safety Architect** handles **Foundational Analysis Of Safety Engineering Requirements**.
81. **Initalization** is an informatical and systemic process.
82. **Initalization** requires **Configuration** and **System Model**.
83. **Initalization** yields **List Of Requirements** and **Llm Client**.
84. **Classifying Requirements** is an informatical and systemic process.
85. **Classifying Requirements** requires **List Of Requirements**, **List Of Subsystems & Functions**, and **Llm Client**.
86. **Classifying Requirements** yields **Requirements Classification Report**.
87. **Identifying Contradictions** is an informatical and systemic process.
88. **Identifying Contradictions** requires **Contradiction Extraction Instructions**, **Llm Client**, and **Requirements Classification Report**.
89. **Identifying Contradictions** yields **Conflicting Requirements Report**.
90. **Identifying Duplications** is an informatical and systemic process.
91. **Identifying Duplications** requires **Duplication Extraction Instructions**, **Llm Client**, and **Requirements Classification Report**.
92. **Identifying Duplications** yields **Duplicate Requirements Report**.
93. **Creating Output Report** is an informatical and systemic process.
94. **Creating Output Report** yields **SAFER Report Set**.
95. **Identifying Coverage Gaps** is an informatical and systemic process.
96. **Identifying Coverage Gaps** requires **List Of Subsystems & Functions** and **Requirements Classification Report**.
97. **Identifying Coverage Gaps** yields **Requirement-to-function Allocation Report**.
98. **Querying** of **Llm Client** is an informatical and systemic process.
99. **Querying** of **Llm Client** requires **Large Language Model**.
100. **Identifying Subsystems** is an informatical and systemic process.
101. **Identifying Subsystems** requires **Subsystem Extraction Instructions**.
102. **Identifying Subsystems** yields **List Of Subsystems & Functions**.

## APPENDIX B. FUNCTION IDENTIFICATION PROMPT

```
1. The <architecture_model> resource is an Object-Process
Methodology (OPM) HTML specification of an architecture
model of a Drone.
2. Identify the system, subsystems, and functions in the
<architecture_model> resource.
3. Provide the results in a json structure with the root
node "results",
    under the root node "results", there is a node for
each one of the primary systems. under each primary
system node is a set of key:value pairs in this pattern:
    {
    "results":{
        "Drone":
            "key1":"subsystem1/function11",
 "key2": "subsystem2/function21"
            },
        "Operator":{
 "key1":"subsystem11/function111",
 "key2": "subsystem12/function121"
            }
        }
    }
   - the key is the alias of the leaf function or a
unique abbreviation of the name if no alias is given.
    - the value is the lineage of the system name/sub-
system name/function name.
    - a system can consist of sub-systems and they are
both OPM things of type "Object".
    - a primary system is a system that has no parent in
the model, i.e., no other system consists of it.
    - a system/ sub-system object can exhibit an OPM
Process, which reprents a function of the owner
system/sub-system.
    - DO NOT REFER to input/output objects that are passed
from function to function as subsystems.
    - look at the  example in the <architecture_example>
but this is just an example and not necessarily a
baseline version.
```

```
    - make sure all the items under "system_functions" are
delineated to a function (an OPM process)

4. Include a placeholder called "Other Function" with the
key "_OF_" which will serve as a catch-all for later
analysis.
```

## APPENDIX C. CLASSIFICATION PROMPT

```
Below is a list of system requirements, under the tag
Drone Safety Requirements.
In the RESOURCES tag, there is ARCHITECTURE resource
which contains a list of the primary systems' functions,
provided as {Alias:Name} pairs.

Write a corresponding System Requirement (SysReq)
according to requirements authoring conventions, in the
correct structure, making sure that it is:
a. A shall statement (The {SYS} shall...)
b. Necessary - essential for achieving system
functionality and regulatory compliance.
c. Clear - expressed simply and concisely to ensure
mutual understanding among stakeholders.
d. Traceable - linked both upward to higher-level
specifications and downward to detailed ones.
e. Verifiable - testable through methods such as
analysis, demonstration, inspection, or testing.
f. Complete - ensuring the system will fully satisfy
customer needs while adhering to standards.
If the requirement is fine as is, keep the original
requirement.

A. Categorize each one of the requirements under ONE of
the Drone's functions, or if your confidence level is
less than 80%, as Other function (_OF_).
The System functions are listed below under the tag
{Function}.
Present only the Function alias for each requirement ID
and use only the aliases in the list of drone functions.

B. Classify each one of the requirements as Functional
(FUNC) or Probabilistic (PROB), or if your confidence
level is less than 80%, as Other type (_OT_).
The types of requirements are listed under the tag
<safety_function_type>
Present only the Type Symbol Type: (FUNC /PROB, /_OT_)
for each ID - do not write the full type name.

C. Provide your confidence level as a number between 0
and 100,
D. Provide an explanation for why you chose a function
E. PRovide an explanation for why you chose a type.

F. Return your results in a JSON structure according to
the following example:
{
    "results":[
        {
 "ReqID": "1_1",
         "System_Requirement": "The engine shall ... "
 "Function": "EN",
 "Type": "PROB",
 "Confidence": 95,
         "Function_Explanation": "The requirement
contains the phrase 'Engine ...'",
 "Type_Explanation": "The requirement contains the phrase
'the probability of ...'"
        },
        {
 "ReqID": "86_0",
         "System_Requirement": "The system shall
provide sufficient control measures."
 "Function": "_OF_",
         "Type": "_OT_",
 "Confidence": 75,
 Function_Explanation":"Text does not match any of the
functions",
 Type_Explanation": "Text does not match any of the
types"
        }
    ]
}
```

## APPENDIX D. CLASSIFIED REQUIREMENTS SAMPLE

| ReqID | Requirements | Function | Type | Confidence | Function_Explanation |
|---|---|---|---|---|---|
| 1000 | The system will fly on high-speed to perform tasks efficiently and cover large areas quickly. | NAV | FUNC | 90 | The requirement refers to the Navigation (NAV) function |
| 1001 | The system will maneuver slowly and precisely to navigate tight spaces or perform delicate operations without risk of accidents. | NAV | FUNC | 85 | The requirement relates to precise maneuvering and navigation in tight spaces |
| 1002 | The system must have a long flight range to cover large areas and perform extended missions without needing frequent recharging or refueling. | EN | PROB | 80 | The requirement is associated with ensuring a long flight range and extended mission capability |
| 1003 | The system should operate with a compact battery to minimize weight and maximize portability | EN | _OT_ | 70 | The requirement pertains to the battery system of the drone |
| 1004 | The autonomous system must have multiple independent communication channels to ensure reliable data transmission and control, preventing failure in case one channel is compromised. | TD | FUNC | 85 | The requirement involves reliable data transmission and control |
| 1005 | The system must include additional backup communication links that activate automatically if the primary communication system fails, ensuring continuous operation and control. | TD | FUNC | 90 | The requirement includes backup communication links for continuous operation |
| 1006 | The autonomous system must be equipped with multiple types of sensors (e.g., radar, lidar, and cameras) to detect and avoid obstacles in real-time. | _OF_ | _OT_ | 75 | The requirement involves sensor integration for obstacle detection |
| 1007 | The system must employ multiple, independent algorithms for obstacle detection and avoidance to ensure robust performance and prevent collisions. | SUP | FUNC | 85 | The requirement relates to obstacle detection and collision prevention |
| 1008 | The system must utilize multiple, independent communication channels to receive data, ensuring that data transmission continues even if one channel fails. | TD | FUNC | 80 | The requirement involves utilizing multiple communication channels for data reception |
| 1009 | The system must have backup data receiving links that automatically activate if the primary link fails to maintain continuous data reception. | TD | PROB | 90 | The requirement includes backup data receiving links for continuous data reception |